\documentclass{llncs}
\usepackage{makeidx}  % allows for indexgeneration
\usepackage{graphicx}
\begin{document}
\mainmatter              % start of the contributions
\title{Argudas: arguing with gene expression information}
\titlerunning{Argudas}  % abbreviated title (for running head)
%                                     also used for the TOC unless
%                                     \toctitle is used
%
\author{Kenneth M$^c$Leod\inst{1} \and Gus Ferguson\inst{1}  \and Albert Burger \inst{1,2}}
\authorrunning{Kenneth M$^c$Leod et al.} % abbreviated author list (for running head)
%
%%%% list of authors for the TOC (use if author list has to be modified)
\tocauthor{Kenneth M$^c$Leod, Gus Ferguson and Albert Burger}
\institute{Heriot-Watt University, Edinburgh, EH14 4AS, UK\\
\email{kenneth.mcleod@hw.ac.uk},\\
\texttt{http://www.macs.hw.ac.uk/bisel}
\and
MRC Human Genetics Unit,
Edinburgh,
EH4 2UX, UK}

\maketitle              % typeset the title of the contribution

\begin{abstract}
\emph{In situ} hybridisation gene expression information helps biologists identify where a gene is expressed.  However, the databases that republish the experimental information are often both incomplete and inconsistent.  This paper examines a system, Argudas, designed to help tackle these issues. Argudas is an evolution of an existing system, and so that system is reviewed as a means of both explaining and justifying the behaviour of Argudas.  Throughout the discussion of Argudas a number of issues will be raised including the appropriateness of argumentation in biology and the challenges faced when integrating apparently similar online biological databases.
\keywords{argumentation, gene expression, information integration}
\end{abstract}

%%%%%%%%%%%%%%%%%%%%%%%%%%%%%%%%%%%%%%%%%%%
%%%%%%%%%%%%%%%%%%%%%%%%%%%%%%%%%%%%%%%%%%%
%%%%%%%%%%%%%%%%%%%%%%%%%%%%%%%%%%%%%%%%%%%
\section{Gene expression, inconsistency, and incompleteness} \label{sec:background}

Gene expression information describes whether or not a gene is expressed (active) in a location.  Broadly speaking there are two types of gene expression information: those that focus on where the gene is expressed, and those whose primary concern is the strength of expression.  This work focuses on the former category, in particular a technology called \emph{in situ} hybridisation gene expression.

Information on gene expression is often given in relation to a tissue\footnote{It may be given for a point in 3D space.} in a particular model organism.  In this work the model organism of interest is the mouse.  This organism is studied from conception until adulthood.  The time window is split into 28 so-called Theiler Stages.  Each stage has its own anatomy, and corresponding anatomy ontology called EMAP \cite{emapBook}.  The first 26 stages cover the developmental mouse: the mouse from conception until birth.  Stage 27 is the new born mouse, and 28 the adult.

The result of an \emph{in situ} experiment is an image displaying an area of a mouse (from a particular Theiler Stage) in which some subsections of the mouse are highly coloured.  Areas of colour indicate that the gene is expressed in that location.  In addition to showing where the gene is expressed, the image provides some indication of the volume (or strength) of expression.  The more intense the colour, the stronger the expression.

Result images are analysed manually by examining the image under a microscope.  A human expert determines in which tissues the gene is expressed, and at what level of expression.  As volume information  is not the main focus of the experiment, its description is often vague using loose natural language terms such as \emph{strong}, \emph{moderate}. \emph{weak} or \emph{present}.  For example, the gene \emph{bmp4} is strongly expressed in the future brain from Theiler Stage 15.  

Once an \emph{in situ} gene expression experiment is completed the experiment may be published in a traditional journal.  Regardless of whether or not this is true, the experiment almost always will be published by one (or more) online resources.  Two of the main resources in the current domain of interest are EMAGE\footnote{http://www.emouseatlas.org/emage} and GXD\footnote{http://www.informatics.jax.org} - both use the EMAP anatomy ontology.  These online databases publish so-called \emph{annotations} that contain particular types of information: provenance, details of the technique used, analysis of the result, and perhaps some indication of how reliable the resource believes the experiment to be.   It is possible to supply information directly to the resources, and thus omit the traditional journal publication.  Often such a route is favoured by the large-scale projects that conduct a large number of experiments.

Although EMAGE and GXD are large resources they cannot be considered complete \cite{me_appliedComputing_salamanca}.  There is a range of reasons for this phenomenon including: some large scale projects publish their own results in a proprietary database, some experiments are deemed of insufficient quality by the resource curators, and others simply `slip under the radar'.  Consequently, in order to build as complete a picture of the domain as possible, it is necessary to consult multiple resources.

In addition to being incomplete, online biological resources are often inconsistent \cite{me_appliedComputing_salamanca}.  In terms of gene expression, this means that the same resource publishes one annotation suggesting the gene is expressed in a particular tissue, and a second annotation suggesting it is not.  As biologists  treat absent and expressed as mutually exclusive, this implies an inconsistency.  Due to the complexity of the underlying experiments there is an array of possible reasons for the different results including: differences in the interpretation of results, unrecognised differences in the experiments, and human error (by either the research team or the resource's curators). As discussed above, it is necessary to use synchronously multiple resources.  Doing so raises the prospect of inconsistency between those resources, in addition to the inconsistency inside each resource. 

Although there are many different methods to tackle the issues described the use of argumentation \cite{bench-capon_dunne_argumentationInAI_07} is considered in this work.  Argumentation will be explored in Section \ref{sec:arg}, before previous work is discussed in Section \ref{sec:previousWork}.  Section \ref{sec:argudas} explores Argudas and  discusses a number of issues currently affecting it before Section \ref{sec:future} outlines possible future work and Section \ref{sec:conclusion} provides a conclusion.

%%%%%%%%%%%%%%%%%%%%%%%%%%%%%%%%%%%%%%%%%%%
%%%%%%%%%%%%%%%%%%%%%%%%%%%%%%%%%%%%%%%%%%%
%%%%%%%%%%%%%%%%%%%%%%%%%%%%%%%%%%%%%%%%%%%
\section{Argumentation} \label{sec:arg}

\emph{Argumentation} \cite{bench-capon_dunne_argumentationInAI_07} is a multidisciplinary field that studies arguments and arguing.  

An \emph{argument} is a reason to believe something is true.  This may be a formal proof, or a piece of natural language: for example, a reason to carry an umbrella.  The crucial attribute of an argument is its defeasibility: an argument may provide a reason to believe something is true, but it does not prove it is definitely true.

Arguing (commonly called \emph{argumentation}) is the process of using arguments to justify a point of view.  This process may take place between multiple agents (human or software) inside a debate, or it may be carried out by a single agent: e.g., a political speech justifying the government's decision to increase taxes.

The subdomain of computational argumentation involves the use of computers for constructing and using arguments.  There is a wide range of domains in which argumentation has been applied including: Artificial Intelligence \& Law \cite{bench-capon_prakken_Argumentation_06}, ontology matching \cite{trojahn_matchingOntologiesWithArg_08}, medical decision support systems \cite{fox_argumentationMedicalSurvey_07},  and agent communication \cite{rahwan_argumentBasedNegotiation_03}.

Argumentation in relation to biology is surprisingly rare.  Jefferys \emph{et al.} \cite{jefferys_argInBio_06} use argumentation to analyse the output of a protein prediction tool. However, most work involves pedagogical efforts to improve the construction of natural language scientific arguments by students, e.g. \cite{bravo_scientificArgInTeachEd_05}.

%%%%%%%%%%%%%%%%%%%%%%%%%%%%%%%%%%%%%%%%%%%
%%%%%%%%%%%%%%%%%%%%%%%%%%%%%%%%%%%%%%%%%%%
%%%%%%%%%%%%%%%%%%%%%%%%%%%%%%%%%%%%%%%%%%%
\section{Arguing over gene expression information} \label{sec:previousWork}

Initial attempts to tackle the problems discussed in Section \ref{sec:background} are documented in length in \cite{me_ismb_08,me_cmna9,sutherland_TaskComposition_09}; a brief reprise will be given here.

M$^{c}$Leod \emph{et al.} \cite{me_ismb_08} describes an early system designed to tackle the above problems.  Essentially, this system allowed a user to enquire if a gene was expressed in a particular tissue from an individual Theiler Stage.  Doing so caused the system to generate a number of arguments, evaluate those arguments, and present the results (argument(s) and associated evaluation) to the user.

For the arguments to be meaningful, they had to be based on expert\footnote{The expert employed for this task was the head curator of EMAGE.} knowledge.  This knowledge was captured in a series of natural language inference rules, so-called \emph{argumentation schemes} \cite{walton_argumentationSchemes_08}.  Essentially, these schemes provide a natural language if-then (\emph{modus ponens}) rule and an associated series of questions that can be asked to ensure the rule's application is suitable in the current context.  Additionally the expert assigned a degree of confidence to each scheme.  

\begin{figure}
\begin{center}
 \includegraphics[width=3.5in]{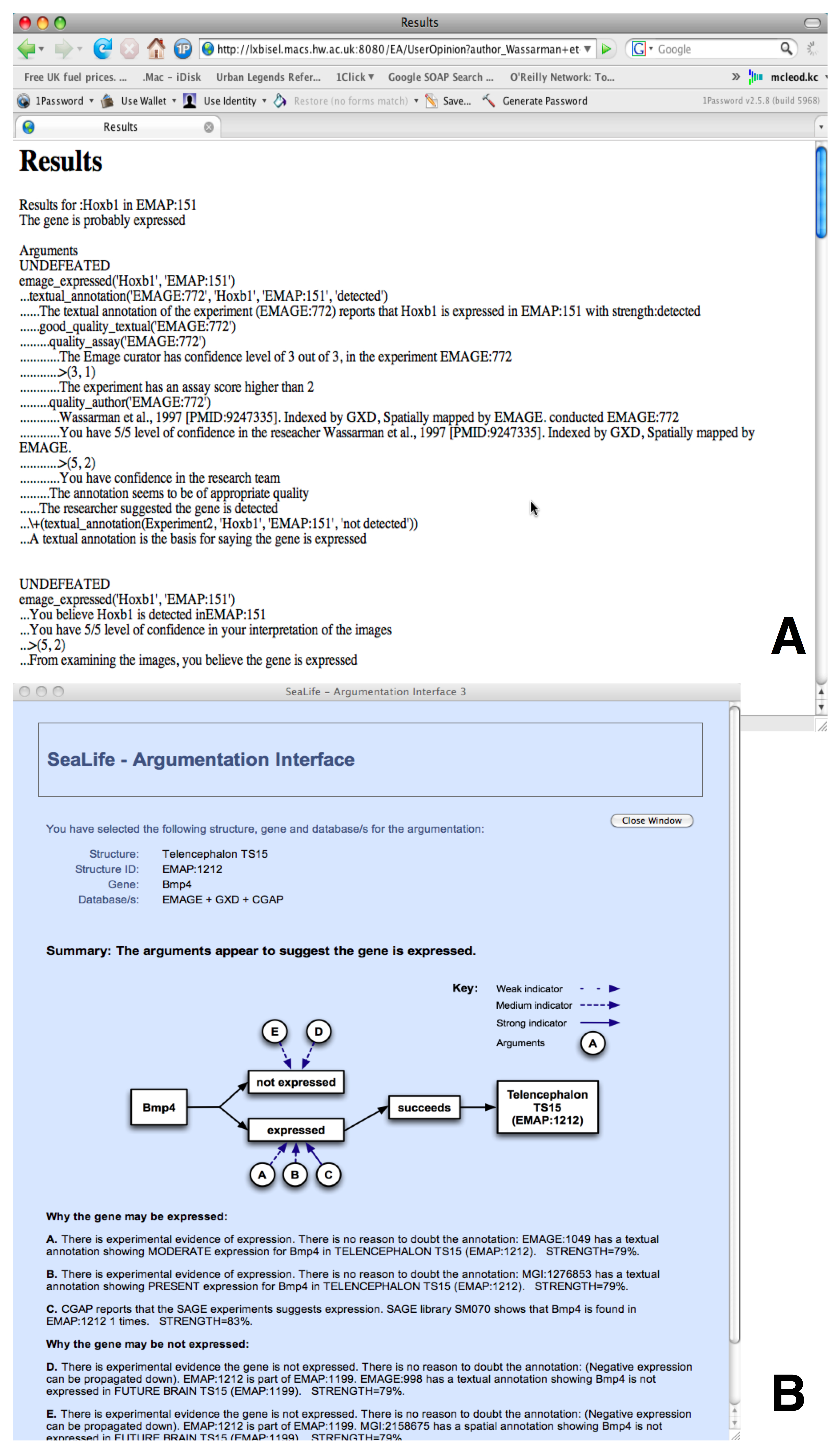} 
\caption{\textbf{A}: The first prototype, results page - two true (undefeated) arguments are displayed using the argumentation engine's default presentation style.
\textbf{B}: The second prototype, results page - a visual summary of the argumentation appears above a list of natural language arguments.}
\label{fig:prot1}
\end{center}
\end{figure}

The natural language schemes were converted into a logical form using the method described by  Verheij \cite{verheij_dialecticalArgumentationWithArgumentationSchemes_03}.  The logic in question is a PROLOG-like logic employed by the ASPIC argumentation engine \cite{fox_argumentationMedicalSurvey_07}.  This tool provides a means to generate arguments, and conduct a virtual debate between two agents in order to determine which argument is stronger. Arguments were created by the ASPIC argumentation engine using the rules, and biological facts (information pulled dynamically at runtime from EMAGE and GXD). In M$^{c}$Leod \emph{et al.} \cite{me_ismb_08} the arguments were converted back into natural language and presented to the user as that is the presentation mechanism the expert deemed most suitable.  Figure \ref{fig:prot1} part \textbf{A} shows a screenshot of the results page: two arguments are displayed using one of ASPIC's in-built presentation mechanisms.  Both arguments are \emph{undefeated}, which means the system believes them to be true.  Unfortunately, the default presentation style uses a mixture of natural language and logic rendering it unsuitable for use with the expected user group.

Subsequent work concentrated on the development, and evaluation of an improved interface (see Figure \ref{fig:prot1} part \textbf{B} for the updated results page).    This time the arguments were presented entirely in natural language, and preceded by an image that summarised the argumentation, which was below a single line conclusion (the gene is expressed). 

Sutherland \emph{et al.} \cite{sutherland_TaskComposition_09} discussed the inclusion of this system in a semantic web browser for the life sciences.

Full details of the evaluation can be found in Ferguson \emph{et al.} \cite{evaluation}; though,  M$^{c}$Leod \emph{et al.} \cite{me_cmna9} published the key findings.  In particular, the notion of subjectivity was raised.  It was clear that each prospective user evaluated had their own approach to interpreting the information contained within EMAGE and GXD.  Accordingly, the schemes produced by the expert were occasionally controversial; likewise the associated degrees of confidence.  This phenomenon is explored in relation to argumentation in the philosophical writings of  Perelman \cite{perelman_newRhetoric_69}.  Perelman introduces the notion of an \emph{audience} to capture the idea that each member of an audience has their own reasoning process, and thus each member of the audience will judge the same argument differently.  This means that there is little point in the system trying to decide whether or not the gene is expressed.  Instead the system must generate arguments for and against the gene being expressed, and allow the user to evaluate these arguments in order to reach their own decision.  In effect, the system should aggregate and evaluate data, presenting the good data to inform the user's decision making process.

%%%%%%%%%%%%%%%%%%%%%%%%%%%%%%%%%%%%%%%%%%%
%%%%%%%%%%%%%%%%%%%%%%%%%%%%%%%%%%%%%%%%%%%
%%%%%%%%%%%%%%%%%%%%%%%%%%%%%%%%%%%%%%%%%%%
\section{Argudas: an evolution} \label{sec:argudas}

In 2009 the BBSRC\footnote{www.bbsrc.ac.uk} provided funding to generate a real world tool to help tackle the issues of inconsistency and incompleteness in relation to \emph{in situ} gene expression data for the developmental mouse - this work is undertaken as part of the \emph{Argudas} project.  Argudas is designed to be an evolution of the work described in Section \ref{sec:previousWork}.  Accordingly, the system's mechanics will not be discussed further.  Instead this section focuses on a number of issues that have affected the work.

%%%%%%%%%%%%%%%%%%%%%%%%%%%%%%%%%%%%%%%%%%%%%%%%%%%%%%%%%%%%%%%
%%%%%%%%%%%%%%%%%%%%%%%%%%%%%%%%%%%%%%%%%%%%%%%%%%%%%%%%%%%%%%%
\subsection{Reducing subjectivity by employing multiple experts} \label{sec:argudas:argudas:experts}

As remarked in Section \ref{sec:previousWork} a number of evaluation subjects disagreed with the expert's schemes and his assignment of degrees of confidence to those schemes.  Argudas did not have the resources to create a new set of schemes as this was a substantial task; nevertheless, it was possible to review the degrees of confidence.  To this end, two experts were asked to review the total list of previously generated schemes and award them a score:
\begin{description}
\item [0] disagree with the scheme;
\item [?] don't know - scheme is very weak and is on the border between being rejected and being classified as a weak scheme;
\item [1] weak scheme, i.e. low confidence;
\item [2] moderate scheme, i.e. medium confidence;
\item [3] good scheme, i.e. high confidence.
\end{description}

In total, the two experts were asked to assign a score to 68 schemes.  The experts completely agreed - that is they gave exactly the same score to 16 schemes.  A further 33 schemes were assigned a similar score.  The notion of \emph{similar} being defined as an adjacent score, i.e. if one expert assigned \textbf{2}, then either a \textbf{1} or a \textbf{3} would be classified as similar.  If the two experts assigned scores that were neither adjacent nor exact matches, they were deemed to disagree - this happened with 19 schemes.

In conclusion, the experts broadly agreed on 72\% of the schemes.   This left  28\% of the schemes for which the disagreement was substantial.   Regrettably, one of the experts emigrated shortly after this exercise was completed and was no longer available to assist in the development of Argudas.  Therefore this disagreement was never resolved, nor was its root cause investigated.

Potentially the source of the disagreement was very interesting, as it was not clear whether the conflict between the experts was caused by a genuine difference of opinion or a difference of interpretation.  As the schemes were written in natural language, the latter is a distinct possibility.  

Two experts means the possibility of two different points of view.  Thus when they both agree, the probability of the degree of confidence being accurate increases.  Nevertheless,  disagreement is beneficial, because through it new insights are discovered.  Intuitively, it seems obvious that if the schemes had been produced by multiple experts the range and diversity of the schemes would have been broader.  Furthermore, if two experts had to agree the natural language used to document the schemes, the number of ambiguous phrases would have been reduced.  %Of course, it might be the case that three experts are better than two, four better than three, etcetera.

Yet working with multiple experts would have caused a number of difficulties.  Expert biologists are often geographically disparate.  This in conjunction with their workload means it may be difficult to bring the experts together.  Furthermore, there is an obvious requirement for a formal resolution process to help dissect and settle differences of opinion.  Finally, it must be acknowledged that not all disagreements can be rectified, and that a mechanism for incorporating differences of opinion must exist.   These issues point to the requirement for a framework that enables biologists to work together in order to generate the schemes.  Lindgren \cite{lindgren_evalOfKnowledgeCaptureFrameworkBasedOnSchemes_10} is developing such a framework for the use case of dementia care; however, it is still at an early stage, and thus cannot be employed here.

%%%%%%%%%%%%%%%%%%%%%%%%%%%%%%%%%%%%%%%%%%%%%%%%%%%%%%%%%%%%%%%
%%%%%%%%%%%%%%%%%%%%%%%%%%%%%%%%%%%%%%%%%%%%%%%%%%%%%%%%%%%%%%%
\subsection{Reducing information overload} \label{sec:argudas:argudas:args}

As Argudas was developed it became clear that the number of arguments generated varied enormously. For some queries there were no annotations and therefore no arguments.  For other queries over ten annotations were retrieved from EMAGE and GXD, accordingly a large number of arguments were generated.  For example, arguing for \emph{bmp4} - future brain in stage 15 generated two hundred and fifteen arguments.  Clearly, no biologist would read all the arguments, hence there can be no guarantee that (s)he would read all the important information.  This realisation led to the conclusion that the potential number of arguments was too high, and steps were taken to reduce it.

Although all of the arguments were unique in terms of their content (wording, and order of words) semantically several arguments seemed to duplicate one another.  Identifying semantically equivalent arguments is not a minor task.  The definition of \emph{equivalent} seems to depend on the individual using the system and the biological task they wish to perform.  

There are a number of common interpretations and actions that are not appropriate for certain biological tasks, and which individual biologists may, in general, reject.  For example, the EMAP anatomy ontology is defined using part-of relationships.  Consequently, positive levels of expression are routinely propagated up the ontology to higher level tissues; for example, if \emph{bmp4} is weakly expressed in the telencephalon, it is normally correct to say that \emph{bmp4} is weakly expressed in the future brain.  Nevertheless, many biologists prefer direct annotations over propagated ones, thus if a second annotation suggested \emph{bmp4} was not detected in the future brain, the second annotation would take precedence.  

Likewise, there is a similar problem with the granularity of information desired.  Finding two distinct annotations with the same conclusion is a powerful argument for trusting the conclusion.  However, the granularity of information desired affects the decision as to whether or not the annotations are in agreement.  Assume there are two annotations: one annotation suggests \emph{bmp4} is strongly expressed in the future brain, and a second annotation demonstrates \emph{bmp4} is weakly expressed in the future brain.  If the biologist is attempting to determine if the gene is expressed or not expressed, then these annotations may be taken to agree.  Yet, if the aim is to determine the level of expression, these annotations are conflicting.

The goal of reducing the number of potential arguments was further hindered by a request for more positive aspects to be highlighted. For instance, although an argument is created when the probe\footnote{A probe is used to detect the presence of an expressed gene.} information is absent for an experiment, no argument is created when it is present.

In summary, Argudas' users appear to wish for a broader range of potential arguments, and yet a smaller number of realised arguments.  Reconciling these competing aims seemed improbable, until it was remarked that the problem was not the volume of the arguments but the amount of text to be read.

%%%%%%%%%%%%%%%%%%%%%%%%%%%%%%%%%%%%%%%%%%%%%%%%%%%%%%%%%%%%%%%
%%%%%%%%%%%%%%%%%%%%%%%%%%%%%%%%%%%%%%%%%%%%%%%%%%%%%%%%%%%%%%%
\subsection{The notion of argument reconsidered} \label{sec:argudas:argudas:new}

Previous work, and an initial version of Argudas, used the ASPIC argumentation engine to generate and evaluate arguments inside a virtual debate.  These arguments were presented to the user as a natural language paragraph - this display mechanism was chosen as it was the preference of the original expert.  However, feedback suggested this choice was subjective \cite{me_cmna9}.  Furthermore, the issues discussed in Section \ref{sec:argudas:argudas:args} appeared to imply that it was sub-optimal.  There was a clear need to find an alternative method for displaying arguments.

During internal discussions it was proposed that the argumentation mechanism should be reconsidered.  This approach was based on the belief that users wanted quick access to certain key attributes of the annotation.   The theory was that there was no need to employ the argumentation engine to create and evaluate arguments.  Instead, the most important schemes (as identified by the process described in Section \ref{sec:argudas:argudas:experts}), should be the basis for a range of key attributes that describe the annotation.  The schemes indicate whether or not the information stored in EMAGE/GXD for a particular annotation should increase or decrease a user's confidence in that annotation. As such, Argudas would extract information from the resources to generate arguments; however, the analysis of those arguments and all reasoning based on those arguments would be left to the user.

\begin{figure}
\begin{center}
 \includegraphics[width=3in]{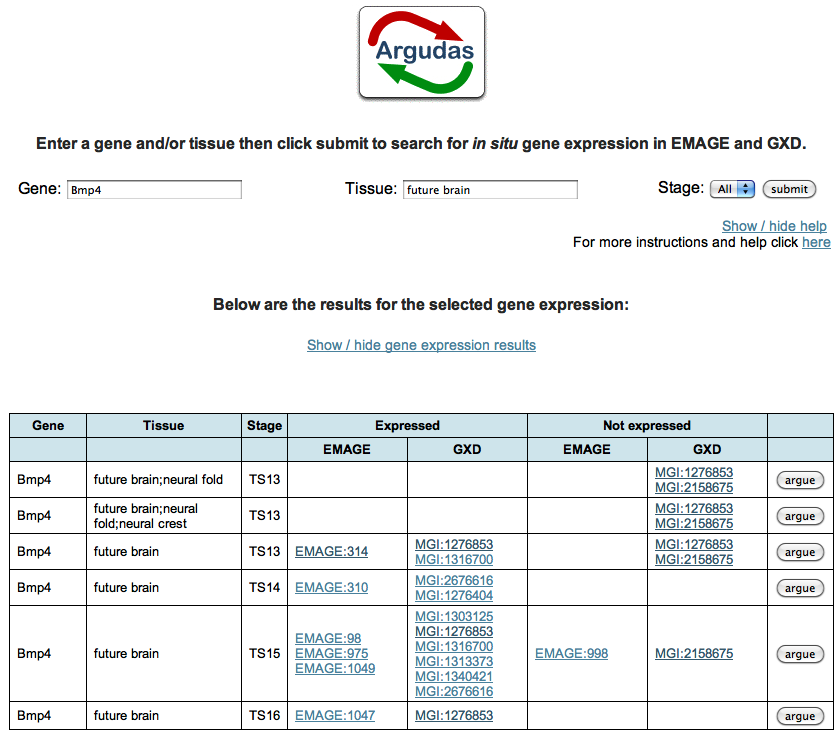} 
\caption{Mock-up of the user interface: simple form allows user to search for gene and/or tissue in relation to a particular Theiler Stage.  Doing so produces a table summarising the relevant annotations found in EMAGE and GXD.}
\label{fig:gui2a}
\end{center}
\end{figure}

In order to test this hypothesis two mock interfaces were created and evaluated.  There were three steps to each interface.  The first two steps were the same: select a gene and/or tissue of interest; report on the available annotations and allow the user to ask for more information if desired.  Figure \ref{fig:gui2a} shows both of these: initially the query is \emph{bmp4} - future brain in all stages; the query causes all combinations of the gene and tissue to be displayed in a table.  The table presents all relevant annotations, summarises what each annotation shows, and provides a link to the resource's web page for that annotation.

In some situations the table in Figure \ref{fig:gui2a} would be enough to resolve a biologist's question; i.e., it is clear that \emph{bmp4} is expressed in the future brain in stage 14.  In the situations where the table is not helpful, or does not provide enough information, clicking the \emph{argue} button provides a range of arguments.

\begin{figure}
\begin{center}
 \includegraphics[width=3in]{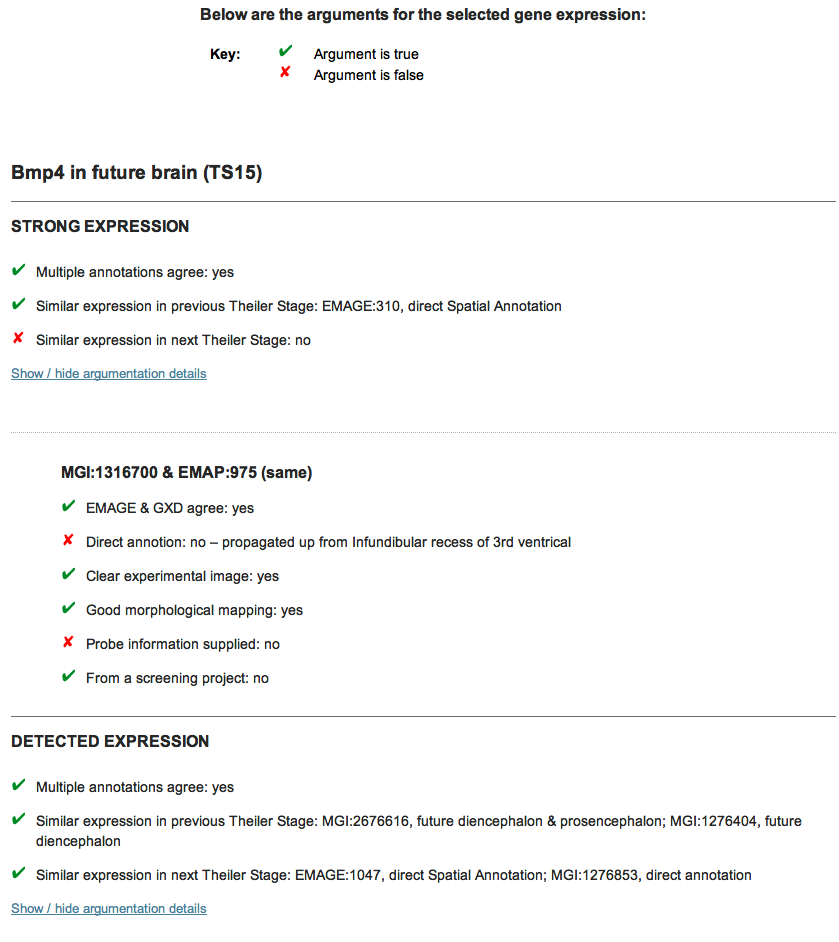} 
\caption{Mock-up of the user interface: potential display mechanism for arguments.  Clicking on an \emph{argue} button from Figure \ref{fig:gui2a} produces an output where different attributes are assigned a positive (tick) or negative (cross) indicator.  Positive indicators demonstrate the annotation is more likely to be correct.}
\label{fig:gui2b}
\end{center}
\end{figure}

In the first mock interface a number of textual arguments were displayed - in a similar manner to Figure \ref{fig:prot1} part \textbf{B}.  The second interface can be seen in Figure \ref{fig:gui2b} - the arguments are now a list of key attributes such as  \emph{multiple annotations agree}.  Whether or not an attribute should strengthen a user's confidence in the annotation is indicated with a tick or cross.  The attributes are divided into two layers - firstly by expression level, and then by annotation.  Thus for each level of expression there are three attributes that indicate how likely that level of expression is.  Asking for more information causes the second layer of attributes to appear.  This allows the user to evaluate the annotations individually, and collectively as a group that promotes a specific expression level.

%%%%%%%%%%%%%%%%%%%%%%%%%%%%%%%%%%%%%%%%%%%%%%%%%%%%%%%%%%%%%%%
\subsubsection{Evaluation} \label{sec:argudas:argudas:args:eval}
The two presentation styles of argument were evaluated with the assistance of two expert users from the Medical Research Council's Human Genetic Unit\footnote{http://hgu.mrc.ac.uk} (HGU). One expert had participated in the development previously, as discussed in Section \ref{sec:argudas:argudas:experts}. The expert evaluations were undertaken independently with no discussion between the experts prior to the evaluation.

Each expert user was presented with a description of the planned evaluation, then a structured walkthrough was conducted. Using a protocol, the user was guided through each interface using the same search example: \emph{bmp4} - future brain - stage 15. They were asked to raise any issues or aspects they liked or disliked while undertaking the interface evaluations. The experts were then asked to score the interfaces out of ten in terms of their usability. In conclusion, a limited set of questions was asked to determine the user's opinions on the requirements for refining aspects of the interface and argument presentations.

Although the evaluation was too limited to allow any statistical analysis, the qualitative data collected provides useful indications for future development of the system. Both experts were in complete agreement with regard to the broad future of Argudas. The experts indicated a preference for the tick/cross style of presenting arguments, and agreed that this presentation style still provided them with too much to read. Both expert users suggested tabulation to address this issue.

The experts differed in their implementation of the tables, with one believing the existing expression level layer of attributes was acceptable, and that only the annotation layer should be converted into a table with one table for each expression layer. The other expert user preferred all the expression level layer attributes in one table, and all the annotation layer attributes in a second table. This leads to the association between an expression level and an annotation being lost,  and potentially results in a very large annotation level table being generated.

Although limited, the evaluation demonstrated that the second interface style, in which arguments become attributes, is preferred over the previous version. This is significant, as it removes the requirement for argument evaluation by the argumentation engine. Lastly, the evaluation illustrated that although the second iteration of the user interface is moving in the right direction, further iteration and refinement is required.

%%%%%%%%%%%%%%%%%%%%%%%%%%%%%%%%%%%%%%%%%%%%%%%%%%%%%%%%%%%%%%%
%%%%%%%%%%%%%%%%%%%%%%%%%%%%%%%%%%%%%%%%%%%%%%%%%%%%%%%%%%%%%%%
\subsection{Extending Argudas for richer argumentation} \label{sec:argudas:argudas:resources}

Argudas aims to improve on previous work with the integration of further resources - more resources means more information and richer arguments.  Initially the microarray data contained in the ArrayExpress\footnote{www.ebi.ac.uk/arrayexpress/} resource was targeted.  Unfortunately, this highlighted a number of  integration issues that could not be resolved in the project's time frame.

Firstly, the ArrayExpress resource does not use the EMAP anatomy ontology.  Secondly, accessing the data held by ArrayExpress was difficult as they did not provide a direct programmatic access to their database.  Instead access was via a RESTFUL web service; that service provided limited functionality, and did not allow access to the data required for this work.  For example, initially\footnote{The team behind the resource were independently working on improving this interface and claimed that such functionality would become available in the future.} it was impossible to ask for all the genes expressed in a healthy mouse's pancreas at stage 24 because ArrayExpress did not compute multi-factor statistics.  That is, they computed which genes were expressed in the pancreas and which genes were expressed in stage 24 separately and there was no way of presenting the intersection at that time.  Finally, ArrayExpress had less data for the developmental mouse than expected: only three stages were covered.  Weighing the costs and benefits it was decided not to pursue this integration further.

As work on ArrayExpress stopped an investigation of the Allen Brain Atlas\footnote{www.brain-map.org/} and GENSAT\footnote{www.gensat.org/} began.  Both of these resources are databases of \emph{in situ} experiments focusing predominately on the adult mouse's nervous system, i.e. brain.  The latter project provides a full database dump.  The former provides an extensive range of RESTFUL interfaces that provide access to the desired information.

However, bringing the data from these two new resources into Argudas is not a simple task.  Neither resource uses the EMAP anatomy - as these resources focus on the brain they have a far finer granularity for the brain tissues than EMAP.  Hence it is necessary to attempt some form of mapping from their respective anatomies to EMAP.  Secondly, these resources use their own measures to describe the level of expression, GENSAT natural language terms and ABA floating point numbers, which also must be mapped across to the corresponding EMAGE/GXD terminology.

Mapping between the different anatomy ontologies employed by the resources is based on a series of alignments produced by Jim\'{e}nez-Lozano \emph{et al.} \cite{argudas_GensatAbaEmageMappings_09}.  As both GENSAT and ABA have a finer granularity than EMAP, mapping from those resources to EMAGE/GXD results in a loss of precision. 

%Anatomy mappings were created by N. Jim\'{e}nez and J. Segura from the National Center for Biotechnology-CSIC, Madrid.  To date these mappings have not been independently verified, thus it should be acknowledged that some of these mappings may not be accurate.  Furthermore, as both GENSAT and ABA have a finer granularity than EMA, mapping from those resources to EMAGE results in a loss of some precision. 

The second task is straightforward for GENSAT as their choice of labels is similar to EMAGE's.  Whereas EMAGE\footnote{EMAGE and GXD use very similar labels for example \emph{absent} instead of \emph{not detected}.} has \emph{not detected}, \emph{detected}, \emph{weak}, \emph{moderate}, and \emph{strong} GENSAT has \emph{not done}, \emph{undetectable}, \emph{weak signal}, and \emph{moderate to strong signal}.  

Mapping EMAGE/GXD expression levels to ABA is a more complex task.  There are three different measures of expression level published by ABA.  Firstly there is the raw experimental information, then there is the average information (across all the experiments for a particular gene and tissue), and finally there is a mathematical aggregation of the expression level and expression density.  For current purposes, the first class of information is most suitable.  Subsequently, the ABA generated expression level mappings must be applied.  These mappings are a series of cut-offs that determine whether the expression level is \emph{not expressed}, \emph{weak}, \emph{moderate} or \emph{strong}.  There are different limits for different parts of the brain.  In order for the limits to be applied to the tissues lower down in the anatomy hierarchy, the limits need to be propagated through the brain in a similar manner to the gene expression information.

Once this work has been undertaken it is necessary to determine what level of integration is appropriate for these resources.  At the simplest level it would be possible to merely report the results contained in ABA and GENSAT.  If either of these resources agreed with an annotation from EMAGE/GXD, it would increase the confidence in that annotation.  Fully integrating ABA and GENSAT would require the generation of key attributes for these resources, which is substantially more work and would necessitate involvement from a resource expert.  In the case of ABA such an approach may not be fruitful;  ABA does not publish all the data it collects, accordingly some of the attributes provided by an expert may be hidden from the public, and thereby Argudas.

Although an interested biologist may raise a number of concerns regarding the anatomy and expression level mappings described above, there is currently no other way of aggregating data between the four resources of interest.  Hence, Argudas will progress along this path, continuing to evaluate and adjust the approach according to the feedback from expert users.

%%%%%%%%%%%%%%%%%%%%%%%%%%%%%%%%%%%%%%%%%%%%%%%%%%%%%%%%%%%%%%%%%%%%%%%%%%%%%%
%%%%%%%%%%%%%%%%%%%%%%%%%%%%%%%%%%%%%%%%%%%%%%%%%%%%%%%%%%%%%%%%%%%%%%%%%%%%%%
%%%%%%%%%%%%%%%%%%%%%%%%%%%%%%%%%%%%%%%%%%%%%%%%%%%%%%%%%%%%%%%%%%%%%%%%%%%%%%
\section{Future work} \label{sec:future}
This work set out to model expert knowledge and use it to reason with information sources available through the Internet.  In the current use case, this appears to be beyond the scope of what a typical end user wishes.  However, the current use case is relatively constrained, and thus contained: it focuses on one kind of gene expression information for one model organism.    Extending the use case to include different types of biological information, for example gene regulatory networks, makes the use case considerably more complex.  As the intricacy of the biological investigation increases, the need for user support likewise increases.  Argumentation is one possible support mechanism.

Another avenue for future work relates to CUBIST, Combing and Uniting Business Intelligence with Semantic Technologies,  an EU FP7 project that aims to combine the essential features of Semantic Technologies, Business Intelligence, and Visual Analytics.  Data  from unstructured and structured sources will be federated within a Business Intelligence enabled triplet store, before visual analysis techniques such as Formal Concept Analysis \cite{andrews_fcaBedrock_10} are applied.  One of the project's three use cases involves the gene expression data described in Section \ref{sec:background}.  Although it is early in the life of the CUBIST project, a semantic  Extract Transform Load \cite{kimball_warehouseETL_04} process that includes computational argumentation may be envisioned.  The inclusion of argumentation may provide an intelligent transformation of data, and a user-friendly explanation of the transformation.

%%%%%%%%%%%%%%%%%%%%%%%%%%%%%%%%%%%%%%%%%%%%%%%%%%%%%%%%%%%%%%%%%%%%%%%%%%%%%%
%%%%%%%%%%%%%%%%%%%%%%%%%%%%%%%%%%%%%%%%%%%%%%%%%%%%%%%%%%%%%%%%%%%%%%%%%%%%%%
%%%%%%%%%%%%%%%%%%%%%%%%%%%%%%%%%%%%%%%%%%%%%%%%%%%%%%%%%%%%%%%%%%%%%%%%%%%%%%
\section{Conclusion} \label{sec:conclusion}
This paper describes the rationale for generating an argumentation tool (Argudas) to tackle the inconsistency and incompleteness found in \emph{in situ} hybridisation gene expression resources for the developmental mouse.  Furthermore, it discusses the development of Argudas highlighting some of the critical problems still outstanding.  

Although the paradigm of argumentation initially seemed promising for this use case, it is clear that biologists are not interested in the full power of argumentation.  To them the ability to generate and automatically evaluate many arguments is not of primary interest.  Nor is the presentation of a conclusion - they wish to make that decision.  As such there is no place in Argudas for the version of argumentation carried out in previous work.  Instead, the concept of an argument as \emph{a reason to believe something} can be used to present key attributes that provide a good indication of whether or not an annotation can be trusted, and via implication whether or not a gene is expressed.

Computational argumentation may not be wholly appropriate for the current use case, yet that does not mean that the technology cannot be applied to other domains within the Life Sciences.  The current use case is restricted in terms of its complexity.  For more elaborate situations, in which data from multiple fields is aggregated for knowledge generation, the user support provided by argumentation may be more valuable.

The effectiveness of computational argumentation in biology hinges on the quality of the domain modelling.  Regardless of the application domain, the effort required to model domain information is significant.  This cost presents a substantial barrier to the successful adoption of computational argumentation within biology, and raises questions over whether argumentation can reach its full potential within this domain.  Yet the same is true for the Semantic Web in general, and as James Hendler and others \cite{wolstencroft_littleSemanticsLongWayInBiology_05} have stated - a little semantics goes a long way.  

To tackle the incompleteness of a single biological resource it is necessary to aggregate data from other sources.  Argudas is trying to expand beyond its initial online sources; yet, doing so introduces a number of challenges.  All the resources featured in this paper are essentially conducting the same task for the same type of information; however, the resources use different anatomy ontologies, terminologies and methods.  Accordingly, integrating data is not a straightforward task.

The issues faced when integrating data across sources are not unique to \emph{in situ} gene expression for the developmental mouse.  They are applicable to all domains which involve experimentation on model organisms.  Currently, there is no adequate solution to these difficulties and users have to accept some limitations.

In conclusion this paper successfully maps out the future for Argudas, and provides lessons for future argumentation in biological domains.

%%%%%%%%%%%%%%%%%%%%%%%%%%%%%%%%%%%%%%%%%%%%%%%%%%%%%%%%%%%%%%%%%%%%%%%%%%%%%%
%%%%%%%%%%%%%%%%%%%%%%%%%%%%%%%%%%%%%%%%%%%%%%%%%%%%%%%%%%%%%%%%%%%%%%%%%%%%%%
%%%%%%%%%%%%%%%%%%%%%%%%%%%%%%%%%%%%%%%%%%%%%%%%%%%%%%%%%%%%%%%%%%%%%%%%%%%%%%
\section*{Acknowledgments}
The support of the MRC's Human Genetics Unit is greatly appreciated, in particular assistance from Dr. S Venkataraman, Dr. I Overton, and Dr. J Christensen.  The authors would also like to thank Dr. L Ng from the Allen Brain Atlas, Dr. N Jim\'{e}nez-Lozano  from the National Center for Biotechnology-CSIC,  and J Segura from Leeds Institute of Molecular Medicine.\\

\noindent
Funding was provided by the  BBSRC project Argudas (BB/G024162/1) and EU project CUBIST (FP7-ICT-2010-257403).

%
% ---- Bibliography ----
%
\bibliographystyle{splncs03}

%\clearpage
%\addtocmark[2]{Author Index} % additional numbered TOC entry
%\renewcommand{\indexname}{Author Index}
%\printindex
%\clearpage
%\addtocmark[2]{Subject Index} % additional numbered TOC entry
%\markboth{Subject Index}{Subject Index}
%\renewcommand{\indexname}{Subject Index}
%\input{subjidx.ind}
\end{document}